\documentstyle[12pt]{article}
\begin{document}

\noindent {\large\bf Existence Theorems for $\frac{\pi}{n}$ Vortex Scattering}
\vspace{1cm}

\noindent K. Arthur $\left.^{\dag}\right.$
and J. Burzlaff $\left.^{\ddag} \right.$\footnote{On leave of absence from the
School of Mathematical Sciences, Dublin City University, Dublin 9, Ireland.}

\noindent $\dag$ School of Mathematical Sciences, Dublin City University, \\
Dublin 9, Ireland.  \\

\noindent $\ddag$ Fachbereich Physik, Universit\"{a}t Kaiserslautern, \\
Erwin Schr\"{o}dinger Strasse, D-67653 Kaiserslautern, Germany. \\

\vspace{2cm}

\noindent {\large \bf Abstract}
The analysis of $90^{\circ}$ vortex-vortex scattering is extended to
$\frac{\pi}{n}$ scattering in all head-on collisions of $n$ vortices
in the Abelian Higgs model.
A Cauchy problem with initial data that describe the scattering
of $n$ vortices is formulated. It is shown that this Cauchy problem
has a unique global finite-energy solution. The symmetry of the
solution and the form of the local analytic solution then show that
$\frac{\pi}{n}$ scattering is realised.

\vspace{5cm}

\noindent KL-TH-95/7

\newpage

In recent years, analytic and numerical studies have uncovered quite
unexpected scattering processes of soliton-like objects in $(2+1)$ and
$(3+1)$ dimensions. The analytic studies were mainly based on the geodesic
approximation \cite{one:1}, whose validity has recently been proved
for vortices \cite{two:2} and monopoles \cite{thr:3},
and were mainly concerned with the scattering of 2 objects. There are,
however, some interesting analytic results for the scattering of more than 2
extended objects \cite{fou:4} \cite{fiv:5}
\cite{six:6}. Among the most interesting
processes is $90^{\circ}$ scattering of 2 vortices in a head-on collision. In
Ref \cite{sev:7}, existence theorems for this process were given, which
act as an underpinning to numerical methods and approximation
techniques. In this Letter
this analysis is extended to $\frac{\pi}{n}$ scattering in all head-on
collisions of $n$ vortices.

The model we study is the Abelian Higgs model
given by the Lagrangian density
\begin{equation}
{\cal L} = \frac{1}{2}(D_{\mu} \Phi)(D^{\mu} \Phi)^{*} -
\frac{1}{4} F_{\mu \nu} F^{\mu \nu} - \frac{\lambda}{8}
(\mid \Phi \mid^{2} - 1)^{2}.
\end{equation}
$\Phi$ is the complex Higgs field,
$D_{\mu}\Phi = \partial_{\mu} \Phi - \imath A_{\mu} \Phi$, $\mu=0,1,2$, is the
covariant derivative and the gauge fields $F_{\mu \nu}$ are defined in terms
of the real gauge potentials $A_{\mu}$ as
$F_{\mu \nu} = \partial_{\mu} A_{\nu} - \partial_{\nu} A_{\mu}$,
$\mu,\nu =0,1,2$. The indices are raised and lowered with the metric tensor
$g=\mbox{diag}(+1,-1,-1)$. The Euler-Lagrange equations are
\begin{equation}
D^{\mu} D_{\mu} \Phi + \frac{\lambda}{2} \Phi ( \mid \Phi \mid^{2} -1 )=0 ,
\:\:\:
\partial_{\mu}F^{\mu \nu} + \frac{\imath}{2} (\Phi^{*} (D^{\nu} \Phi) -
\Phi(D^{\nu} \Phi)^{*} ) = 0.
\label{el}
\end{equation}

For all $\lambda$ the Euler-Lagrange equations have static, finite energy
$n$-vortex solutions of the form \cite{eig:8},
\begin{equation}
A_{i}(r,\theta) = -\epsilon_{ij} x^{j} n \frac{a(r)}{r^{2}} , \,\, A_{0}=0,
\:\:\:
\Phi(r,\theta) = f(r) \exp [\imath n\theta ]
\label{sln}
\end{equation}
for $i,j=1,2$, where
\begin{equation}
r \frac{\partial }{\partial r} \left(
\frac{1}{r} \frac{\partial a}{\partial r} \right) -
f^{2}(r) [ a(r) - 1] = 0,
\end{equation}
\begin{equation}
2 r \frac{\partial }{\partial r} \left(
r \frac{\partial f}{\partial r} \right) - 2 n^{2} f(r) [a(r) - 1]^{2} -
\lambda r^{2} f(r) [f^{2}(r) - 1] = 0,
\end{equation}
with
\begin{equation}
f(0) = 0 , \:\:\: a(0) = 0, \:\:\:
\lim_{r \rightarrow \infty} f(r) =
\lim_{r \rightarrow \infty} a(r) = 1.
\end{equation}
In the special case $\lambda = 1$, it can be shown \cite{nin:9} that the
solutions actually
satisfy the first order Bogomolnyi equations, so $f$ and $a$ satisfy,
\begin{equation}
r f'- n(1-a)f = \frac{2n}{r} a' + f^{2} -1 = 0.
\end{equation}
In this case, there also exists a $2n$ parameter family of static $n$-vortex
solutions describing vortices located at arbitrary positions. The reason for
its existence is the fact that for $\lambda = 1$, the net force between static
vortices is zero.

We now study the scattering of vortices during the time from shortly before to
shortly after the collision. The fields are taken to be of the form,
\begin{equation}
\Phi(t,\vec{x}) = \hat{\Phi}(\vec{x}) + t \zeta(\vec{x}) , \:\:\:
A_{0}(t,\vec{x}) = 0, \;\;\;
A_{i}(t,\vec{x}) = \hat{A}_{i}(\vec{x}) + t B_{i}(\vec{x}).
\end{equation}
Here $(\hat{\Phi},\hat{A_{i}})$ is taken to be the solution (\ref{sln}). It is
assumed that equations (\ref{el}) can be linearized in $(t \zeta,t B_{i})$
for $t \in (-\epsilon,\epsilon)$, $\epsilon \ll 1$.
These linearized equations are solved for
\[
\zeta = n f(r) k(r)
\]
\begin{equation}
(B_{1},B_{2}) = (-n \sin[(n-1)\theta] \frac{[rk'(r) + nk(r)]}{r},
-n \cos[(n-1)\theta] \frac{[rk'(r) + nk(r)]}{r})
\label{28}
\end{equation}
where $k$ is the solution of the equation,
\begin{equation}
r^{2} k''(r) + rk'(r) - k(r)[n^{2} + r^{2}f^{2}(r)] = 0,
\label{keq}
\end{equation}
with $k(r)e^{r} \rightarrow 1$ as $r \rightarrow \infty$. (The functions
(\ref{28}) have been used by Weinberg \cite{ten:10} in his study of the zero
modes of the static solutions.)

We use the zeroes
of $|\Phi|^{2}$ to define the positions of the vortices. For the
solution (\ref{28}) one obtains
\[
\mid \Phi \mid^{2} = f^{2}(1+2ntk \cos(n \theta) + n^{2} t^{2} k^{2} )
\]
\[
\geq f^{2}(1-n \mid t \mid k)^{2}
\]
For $t \neq 0$, $\mid \Phi \mid^{2}$ has exaclty $n$ zeroes, namely
at $r=\rho$ and $\theta = 0,\frac{2 \pi}{n},\dots,\frac{2 (n-1) \pi}{n}$
for $t < 0$, and at $r=\rho$ and
$\theta = \frac{\pi}{n},\frac{3 \pi}{n},\dots,\frac{(2 (n-1)+1) \pi}{n}$
for $t > 0$. This shows that the solution (\ref{28}) describes
$\frac{\pi}{n}$ scattering. Here $\rho$ is the point where
$k(\rho)=\frac{1}{n \mid t \mid}$. That exactly one such $\rho$ exists
follows from the fact that $k(r)$ is monotonic decreasing, and
behaves like $k_{-n} r^{-n}$ for small r.

The problem with this approach is that the linearization, though
plausible for times shortly before and shortly after the collision, has not
been justified in an entirely rigorous fashion. Guided in our choice of initial
data by the results just discussed we will now bring our discussion to
mathematically
rigorous conclusions.

The first rigorous result concerns the global existence of the solution.
We can now show that for certain initial data a unique global
finite-energy time-dependent solution of equations (\ref{el}) exists in the
Lorentz gauge, $\partial_{\mu} A^{\mu} = 0$. To do this, we first
subtract a backround field $(\hat{\Phi},\hat{A_{\mu}})$ and write,
\begin{equation}
\Phi (t,\vec{x}) = \hat{\Phi} (\vec{x}) + \phi (t,\vec{x}), \:\:\:
A_{\mu} (t,\vec{x}) = \hat{A_{\mu}} (\vec{x}) + a_{\mu} (t,\vec{x}).
\end{equation}
For the backround field we choose the static solution (\ref{sln}). As initial
data we choose:
\[
\phi(0,\vec{x})=0, \:\:\:
a_{0}(0,\vec{x})=0, \:\:\:
a_{i}(0,\vec{x})=0, \:\:\: \mbox{ for } i=1,2;
\]
\begin{equation}
\partial_{t} \phi(0,\vec{x}) = n f(r) k(r), \:\:\:
\partial_{t} a_{0}(0,\vec{x}) = 0,
\label{id}
\end{equation}
\[
\partial_{t} a_{1}(0,\vec{x}) = -n \sin[(n-1)\theta]\frac{[rk'(r) + nk(r)]}{r},
\]
\[
\partial_{t} a_{2}(0,\vec{x}) = -n \cos[(n-1)\theta]\frac{[rk'(r) + nk(r)]}{r}.
\]
For this choice of backround field and initial data, a unique global
finite-energy solution exists, since all the conditions from
Ref \cite{ele:11} are satisfied.

An essential element of the proof which is based on Segal's existence and
uniqueness theorem \cite{twe:12}, is an iteration method. The method
starts with writing the Cauchy problem in the form,
\begin{equation}
\partial_{t} \Psi = -i \tilde{A} \Psi + J
\label{cp}
\end{equation}
where
$\Psi^{t}=(a_{0},p_{0},a_{1},p_{1},a_{2},p_{2},\phi,\pi^{*}),
\; p_{\mu}=\partial_{0} a_{\mu},
\; \pi^{*}=\partial_{0} \phi-\imath a_{0} \phi,$
and where the operator $\tilde{A}$ is defined by,
\begin{equation}
\tilde{A} = \left(
\begin{array}{cccc}
\Gamma & 0 & 0 & 0 \\
0 & \Gamma & 0 & 0 \\
0 & 0 & \Gamma & 0 \\
0 & 0 & 0 & \Gamma
\end{array} \right) \: \: \: , \:
\Gamma = \left(
\begin{array}{cc}
0 & 1 \\
\Delta - m^{2} & 0
\end{array} \right) .
\end{equation}
The vector $J$ can be calculated from equations (\ref{el}) ( and is given in
\cite{sev:7} ).

The solution of the Cauchy problem (\ref{id},\ref{cp}) can be obtained as the
solution of the integro-differential equation,
\begin{equation}
\Psi (t,\vec{x}) = e^{-\imath \tilde{A} t} \Psi(0,\vec{x}) +
\int^{t}_{0} ds \exp[-\imath \tilde{A}(t-s)] J(\Psi(s,\vec{x})).
\end{equation}
In turn this integro-differential equation is solved by the limit of a
sequence of successive approximations $\Psi_{n}$ defined by the formula:
\begin{equation}
\Psi_{n+1} (t,\vec{x}) = e^{-\imath \tilde{A} t} \Psi(0,\vec{x}) +
\int^{t}_{0} ds \exp[-\imath \tilde{A}(t-s)] J(\Psi_{n}(s,\vec{x})),
\label{sa}
\end{equation}
where $\Psi_{0} =
\Psi (0,\vec{x})$, with the
initial data (\ref{id}). We now establish certain symmetries of the initial
data $\Psi_{0}$ and use (\ref{sa}) to establish these symmetries for the
successive approximations $\Psi_{n}$, and finally for the solution of
(\ref{cp}).

The first transformation we study is
$\vec{x} \rightarrow \vec{x}' = S\vec{x}$
where $S$ is the orthogonal matrix
\begin{equation}
S = \left(
\begin{array}{cc}
\cos \frac{2 \pi}{n} & - \sin \frac{2 \pi}{n} \\
\sin \frac{2 \pi}{n} & \cos \frac{2 \pi}{n}
\end{array}
\right)
\end{equation}
Under this transformation the initial data change as follows:
\[
\Psi (0,\vec{x}') = M_{1} \Psi (0,\vec{x}),
\]
with
\begin{equation}
M_{1} = \left(
\begin{array}{cccc}
I & 0 & 0 & 0  \\
0 & A & -B & 0  \\
0 & B & A & 0   \\
0 & 0 & 0 & I
\end{array}
\right) \: \: , \: \:
I = \left(
\begin{array}{cc}
1 & 0 \\
0 & 1
\end{array}
\right) \: \: ,
A = \cos \frac{2 \pi}{n} I \:,\:
B = \sin \frac{2 \pi}{n} I .
\end{equation}
We can see that $J(M_{1} \Psi(s,\vec{x})) = M_{1} J(\Psi(s,\vec{x}))$,
$[M_{1},\tilde{A}]=0$ and
\[
\exp( - \imath \tilde{A} t)M_{1}\Psi_{n}(s,\vec{x})
= M_{1} \exp( - \imath \tilde{A} t) \Psi_{n}(s,\vec{x}),
\]
which implies that $\Psi_{n}(t,\vec{x}') = M_{1} \Psi_{n} (t,\vec{x})$ for
all $n \in {\cal N}$. From this follows $\Psi(t,\vec{x}') =
M_{1}\Psi(t,\vec{x})$ for the solution $\Psi$.

Next we study the reflection $(x_{1},x_{2}) \rightarrow (x_{1},-x_{2})$. Under
this transformation the initial data change as follows:
$\psi (t,x_{1},-x_{2}) = M_{2} \psi (t,x_{1},x_{2})$, where
\begin{equation}
M_{2} = \left(
\begin{array}{cccc}
-I & 0 & 0 & 0  \\
0 & -I & 0 & 0  \\
0 & 0 &  I & 0   \\
0 & 0 & 0  & C
\end{array}
\right) \: \: , \: \:
CV = V^{*}
\end{equation}
Furthermore,
$J(M_{2} \Psi(s,\vec{x})) = M_{2} J(\Psi(s,\vec{x}))$.
Again we have $[M_{2},\tilde{A}] = 0$, and
$\Psi_{n}(t,x_{1},-x_{2}) = M_{2} \Psi_{n}(t,x_{1},x_{2})$. From this follows
$\Psi(t,x_{1},-x_{2}) = M_{2} \Psi_{n}(t,x_{1},x_{2})$, for the solution
$\Psi$.

Under the transformations considered, all terms in the energy density:
\begin{equation}
{\cal E} =
\frac{1}{2} \mid D_{0} \Phi \mid^{2} +
\frac{1}{2} \mid D_{i} \Phi \mid^{2} +
\frac{1}{4} F_{ij}^{2} +
\frac{1}{2} F_{0i}^{2} +
\frac{\lambda}{8} ( \mid \Phi \mid^{2} - 1)^{2},
\end{equation}
are invariant. This leads to the following conclusion: If by using functions
like $\mid \Phi \mid^{2}$,$F_{ij}^{2}$ or ${\cal E}$, there is a way of
defining the positions $(x_{1}^{a}(t),x_{2}^{a}(t)),a=1,2$, of exactly $n$
separate vortices, these $n$ positions must lie on $n$ radial lines separated
by an angle $\frac{2 \pi}{n}$ with equal distance from the origin.
(Below we will use the the minima of $\mid \Phi \mid^{2}$
to define these positions.)
Furthermore, one of these radial lines must be the positive $x_{1}$-axis,
or make an angle $\frac{\pi}{n}$ with the positive $x_{1}$-axis. Any vortex
that does not satisfy these conditions immediately leads to $2n-1$ other
vortices, because of the symmetries of our solution. Since our
solution is continuous, these positions will change continuously such that
at $t=0$ the $n$ positions coincide, and after the collision the vortices
move again on the radial lines just described. Therefore, they can either go
back on the radial lines they came in on, or go back on radial lines shifted
by an angle $\frac{\pi}{n}$. We will study a further symmetry
and use the Cauchy-Kowalewskyi theorem
\cite{thi:13} to show that the second case is realised.

The last transformation we study is
$\vec{x} \rightarrow M\vec{x}$ where $M$ is the orthogonal matrix
\begin{equation}
M = \left(
\begin{array}{cc}
\cos \frac{\pi}{n} & - \sin \frac{\pi}{n} \\
\sin \frac{\pi}{n} & \cos \frac{\pi}{n}
\end{array}
\right).
\end{equation}
Under this transformation the initial data change as follows:
\[
\Psi (0,M\vec{x}) = M_{3} \Psi (0,\vec{x}),
\]
with
\begin{equation}
M_{3} = \left(
\begin{array}{cccc}
-\sigma & 0 & 0 & 0 \\
0 & C & -D & 0 \\
0 & D & C & 0 \\
0 & 0 & 0 & -\sigma
\end{array}
\right) \: \: , \: \:
\sigma = \left(
\begin{array}{cc}
1 & 0 \\
0 & -1
\end{array}
\right) \: \: ,
C = \cos \frac{\pi}{n} \sigma \:,\:
D = \sin \frac{\pi}{n} \sigma .
\end{equation}
We can see that $J(M_{3} \Psi (s,\vec{x})) = -M_{3} J(\Psi(s,\vec{x}))$,
$\{ M_{3},\tilde{A} \} = 0$ and
\[
\exp (\imath \tilde{A} t) M_{3} = M_{3} \exp (-\imath \tilde{A} t).
\]
 From this follows $\Psi (-t,M\vec{x}) = M_{3} \Psi (t,\vec{x})$,
and we see that all terms in the energy density are invariant
under the transformation $(t,\vec{x}) \rightarrow (-t,M\vec{x})$.
This establishes $\frac{\pi}{n}$ scattering for $n$ vortices.
To show that, for small time $t$, $|\Phi|$ has exactly $n$
minima (so that we can identify $n$ vortices), we use the
Cauchy-Kowalewskyi theorem \cite{thi:13} .

 From Ref \cite{eig:8}, we know that $f$ starts with an $r^{n}$ term and
$a$ starts with an $r^{2}$ term. Equation (\ref{keq}) shows that asymptotically
near $r=0$,
\begin{equation}
k \sim k_{-n} r^{-n} + k_{n} r^{n}.
\end{equation}
Using the same techniques as in Ref \cite{sev:7}, we can now establish the
analyticity of the initial data, and verify that all the conditions of the
Cauchy-Kowalewskyi theorem \cite{thi:13} are satisfied. We therefore have
an analytic solution near the origin. This solution leads to the following
asymptotic expression for $\mid \Phi \mid^{2}$:
\begin{equation}
|\Phi|^{2} (t,\vec{x}) \sim f_{n}^{2} (x_{1}^{2} + x_{2}^{2})^{n} +
2ntf_{n}^{2} k_{-n} \sum_{p=0}^{[n/2]} (-1)^{n+p}
\left(
\begin{array}{c}
n \\ 2p
\end{array}
\right)
x_{1}^{n-2p} x_{2}^{2p} + Kt^2,
\label{app}
\end{equation}
where we have assumed that $x_1$ and $x_2$ are of order $\epsilon$,
and $t$ is of order $\epsilon^{n}$, so that higher order terms can
be ignored. ( $K$ is a constant that can be expressed in terms of
the coefficients of the expansions of $a$, $f$, and $k$.
$[n/2]$ is the highest integer that does not exceed $n/2$.) For
$t \neq 0$, the approximation (\ref{app}) of $|\Phi|^{2}$ has
exactly $n$ minima which lie at $r=(n|t|k_{-n})^{1/n}$
and $\theta = 0, \frac{2\pi}{n}, ..., \frac{2(n-1)\pi}{n}$
for $t<0$, and at $r=(n|t|k_{-n})^{1/n}$
and $\theta = \frac{\pi}{n}, \frac{3\pi}{n},
..., \frac{(2(n-1)+1)\pi}{n}$ for $t>0$. This establishes the
$\frac{\pi}{n}$ symmetry of the process. Our rigorous method
is, however, not capable of following the minima for large
time $t$.

{\large Acknowledgement}

\noindent This work was supported in part by the GKSS-Forschungszentrum
Geesthacht.


\newpage
\begin{thebibliography}{99}

\bibitem {one:1}
        Manton, N.S., \,
        Phys. Lett. B {\bf 110}, 54 (1982).

\bibitem {two:2}
        Stuart, D., \,
        Comm. Math. Phys. {\bf 159}, 51 (1994).

\bibitem {thr:3}
        Stuart, D., \,
        {\em The geodesic approximation for Yang-Mills-Higgs
        equations},\,
        U.C. Davis preprint (1994).

\bibitem {fou:4}
        Kudryavtsev, A., Piette, B., and Zakrzewski, W.J., \,
        Phys. Lett. A {\bf 180}, 119 (1993).

\bibitem {fiv:5}
      Dziarmaga, J., \,
      Phys. Rev. D {\bf 49}, 5609 (1994).

\bibitem {six:6}
        Manton, N.S., and Murray, M.K., \,
        {\em Symmetric monopoles},\,
        University of Cambridge preprint (1994).

\bibitem {sev:7}
        Abdelwahid, F., and Burzlaff, J., \,
        J. Math. Phys. {\bf 35}, 4651 (1994).

\bibitem {eig:8}
        Plohr, B.J., \,
        Doctoral Dissertation, Princeton University (1980); \\
        J. Math. Phys. {\bf 22}, 2184 (1981).

\bibitem {nin:9}
        Jaffe, A., and Taubes, C., \,
        {\em Vortices and Monopoles},\,
        Birkh\"{a}user, Boston, 1980.

\bibitem {ten:10}
        Weinberg, E.J., \,
        Phys. Rev. D {\bf 19}, 3008 (1979).

\bibitem {ele:11}
        Burzlaff, J., and Moncrief, V., \,
        J. Math. Phys. {\bf 26}, 1368 (1985).

\bibitem {twe:12}
        Segal, I., \,
        Ann. Math. {\bf 78}, 339 (1963).

\bibitem {thi:13}
        See, e.g., Petrovsky, I.G., \,
        {\em Lectures on Partial Differential Equations},\,
        Interscience, New York, 1954.

\end{thebibliography}
\end{document}